\documentclass[conference]{IEEEtran}
\usepackage{cite}
\usepackage{amsmath,amssymb,amsfonts}
\usepackage{algorithmic}
\usepackage{graphicx}
\usepackage{textcomp}
\usepackage{xcolor}
\usepackage{color, colortbl}
\usepackage{url}
\usepackage{subfig}
\def\BibTeX{{\rm B\kern-.05em{\sc i\kern-.025em b}\kern-.08em
    T\kern-.1667em\lower.7ex\hbox{E}\kern-.125emX}}

\usepackage{alltt                                    % I like these
	, multirow
	,subfig
	, booktabs
	, listings
	, graphicx
	,float
	,cite
	,verbatim
	,mathtools
	,url
}
\usepackage{tikz}
\usetikzlibrary{patterns}
\usepackage{pgfplots}
\usepackage{pgf-pie}

\newif\ifpienumberinlegend
\pgfkeys{/number in legend/.code=
    \expandafter\let\expandafter\ifpienumberinlegend
    \csname if#1\endcsname
    \ifpienumberinlegend

    \def\beforenumber##1\afternumber{}%
    \fi,
    /number in legend/.default=true
}

\usepackage{amsmath} % Required for \boldsymbol
\usepackage{longtable}
\usepackage{url}
\usepackage{pgf-pie}
\usepackage{footnote}
\usepackage{enumitem}
\usepackage{siunitx}
\usepackage{ifthen}    % For conditional statements
\usepackage{soul}      % For striking out text
\usepackage{float}
\usepackage{lipsum}
\usepackage{mwe}
\usepackage{wrapfig}
\usepackage{tcolorbox}
\usepackage{subcaption}

\newboolean{showcomments}
\setboolean{showcomments}{true} % Change to false to hide comments
\ifthenelse{\boolean{showcomments}}
{% If showcomments is true
  \newcommand{\nbc}[3]{
    \colorbox{#3}{\bfseries\sffamily\scriptsize\textcolor{white}{#1}}
    {\textcolor{#3}{\sf\small$\blacktriangleright$\textit{#2}$\blacktriangleleft$}}
  }
}
{% If showcomments is false
  \newcommand{\nbc}[3]{}
}

\begin{document}

% \title{Breaking Barriers: Analyzing Female Leadership and Expertise in Software Systems Research}
% \title{Do Gender Representations Influence Contributions and Collaborations in Software Systems Research? An Exploratory Study}
\title{Gender Disparities in Contributions, Leadership, and Collaboration: An Exploratory Study on Software Systems Research}

\author{Shamse Tasnim Cynthia \hspace{4mm} Saikat Mondal \hspace{4mm} Joy Krishan Das \hspace{4mm}  Banani Roy\\
\normalsize Department of Computer Science, University of Saskatchewan, Canada\\
\normalsize \{shamse.cynthia, saikat.mondal, joy.das,  banani.roy\}@usask.ca
}

\maketitle

\begin{abstract}
Gender diversity enhances research by bringing diverse perspectives and innovative approaches. It ensures equitable solutions that address the needs of diverse populations. However, gender disparity persists in research where women remain underrepresented, which might limit diversity and innovation. Many even leave scientific careers as their contributions often go unnoticed and undervalued. Therefore, understanding gender-based contributions and collaboration dynamics is crucial to addressing this gap and creating a more inclusive research environment.
In this study, we analyzed 2,000 articles published over the past decade in the Journal of Systems and Software (JSS). From these, we selected 384 articles that detailed authors' contributions and contained both female and male authors to investigate gender-based contributions.
Our contributions are fourfold.
First, we analyzed women's engagement in software systems research. Our analysis showed that only 32.74\% of the total authors are women and female-led or supervised studies were fewer than those of men. 
Second, we investigated female authors' contributions across 14 major roles. Interestingly, we found that women contributed comparably to men in most roles, with more contributions in conceptualization, writing, and reviewing articles.
Third, we explored the areas of software systems research and found that female authors are more actively involved in human-centric research domains.
Finally, we analyzed gender-based collaboration dynamics. Our findings revealed that female supervisors tended to collaborate locally more often than national-level collaborations.
Our study highlights that females' contributions to software systems research are comparable to those of men. Therefore, the barriers need to be addressed to enhance female participation and ensure equity and inclusivity in research.

\end{abstract}


% --------------------------------
\section{Introduction}
% --------------------------------
Diversity boosts innovation when actively maintained and protected from stereotypes \cite{albusays2021diversity, blincoe2019perceptions, happe2021frustrations}. Gender diversity enhances efficiency and fosters ideas for robust, user-centered systems \cite{vasilescu2015gender, earley2000creating, jehn1999differences}. 
However, significant gender disparity in research persists, with women underrepresented, which might limit the usability and functionality of complex systems \cite{hyde2014gender, bastarrica2018affirmative}.
For example, UNESCO reported in 2023 that as of 2020, women made up only 31\% of researchers in science and engineering \cite{bello2021smart}, with their contributions primarily confined to specific regions (e.g., European countries) \cite{felizardo2021global}.
This gender disparity in  
research contributes to biased software design \cite{spangenberg2021holistic}, negative gender stereotypes \cite{cundiff2016gender}, and reduced mentorship opportunities~\cite{hyrynsalmi2019underrepresentation}.

Several studies have examined female participation in the software industry \cite{trinkenreich2022women, kazmi2014women, maheshwari2023review} and software research \cite{cavero2015evolution, boekhout2021gender, bano2019gender, moldovan2024diversity, catolino2019gender}. 
For instance, Cavero et al. \cite{catolino2019gender} reported an annual 3.5\% growth in women's participation from 1960 to 2010. However, they focused mainly on the human-computer interaction (HCI) field. Mathew et al. \cite{mathew2018finding} highlighted women's underrepresentation in software engineering articles. Studies also show that many women leave scientific careers due to the lack of recognition for their contributions \cite{chang2021job, esteves2017crafting}. Therefore, despite several studies on software research focusing primarily on participation statistics, an in-depth analysis of women's specific contributions, leadership, and collaboration dynamics is warranted.

In this study, we conducted a gender-based exploration of software systems research, focusing on four key aspects: a) women's participation and leadership in the field, b) their contributions and level of involvement in research work, c) their specialization and interests in specific areas of software systems, and d) their collaboration dynamics at local, national, and international levels. We addressed four research questions and presented four key findings.

\textbf{RQ\textsubscript{1}}(\textit{participation and leadership})\textbf{: What percentage of software systems articles have female author(s), and how often do they lead and supervise these articles?} 
We aim to evaluate the prevalence of female authorship and leadership roles (e.g., as lead authors and supervisors), which is crucial for understanding the overall gender representation in software systems research.

\textbf{RQ\textsubscript{2}}(\textit{commitment and contribution}): \textbf{What are the most common contributions by female authors to various aspects of research projects in software systems?} We aim to explore the key contributions of female authors to better highlight their roles, influence, and impact within the field of software systems research.

\textbf{RQ\textsubscript{3}}\textit{(specialization and interest):} \textbf{Which key areas of software systems research showcase the most prominent female leadership and contributions?} 
We examine subdomains with strong female leadership and contributions to provide insight into areas where women are excelling in software systems research.

\textbf{RQ\textsubscript{4}}\textit{(collaboration dynamics)}: \textbf{Do female authors in software systems research collaborate more within their organization, nationally or internationally?} We aim to understand the collaboration patterns of female authors, which is crucial to reveal their networking reach and engagement levels in software system research (SSR).

\noindent\textbf{Replication Package} is available in our online appendix \cite{replication_package}.

% % ---------------------------------------------
% \section{CRedit Details} \label{sec:background}
% % ---------------------------------------------
% \saikat{For the sake of space, we will remove this section and refer to the credit details table from the methodology.}
% % \subsection{CRedit details}
% To analyze the gender-based contribution, we focused on the Journal of Systems and Software (JSS) studies that included the \textit{CRedit authorship contribution statement} section (or “CRedit section” for short). The Contributor Role Taxonomy (CRedit) was developed after realizing that simple-ordered author lists do not show the different types of contributions researchers make \cite{CRedit, allen2019can}. Therefore, a 14-role taxonomy was introduced to represent contributions more accurately, and it has been in use since 2015 \cite{niso2022ansi}. Table~\ref{table:contribution-details} summarizes the CRedit role taxonomy. 

% --------------------------------
\section{Related Work}
% --------------------------------

Several studies have been conducted to analyze diversity in the context of software engineering research \cite{narayanan2023diversity, cavero2015evolution, boekhout2021gender, bano2019gender, felizardo2021global, santana2021scientific, moldovan2024diversity, hosseini2021gender, rodriguez2021perceived}. 
Studies reveal a gradual increase in women's participation in computing research, though disparities persist. 
For instance, Cavero et al. \cite{cavero2015evolution} reported a 3.5\% annual growth in women's involvement in computing research from 1960 to 2010, with significant activity in Human-Computer Interaction. 
Boekhout et al. \cite{boekhout2021gender} highlighted an increase in women starting research careers (from 33\% in 2000 to 40\% recently), though men still produce 15-20\% more publications and dominate senior authorship roles. 
Felizardo et al. \cite{felizardo2021global} noted a global rise in women's research contributions to software engineering, primarily concentrated in European countries.
However, Hosseini et al. \cite{hosseini2021gender} observed fewer publications, citations, and international collaborations for female researchers at Dublin City University from 2013–2018. 
Frachtenberg and Kaner \cite{frachtenberg2022underrepresentation} found women comprised only 10\% of researchers in computer systems, based on 53 peer-reviewed conferences from 2017.
Similarly, Mathew et al. \cite{mathew2018finding} found that women are underrepresented in the top-most cited papers in the software engineering field.

Moreover, the lack of diversity in AI research remains a pressing concern. Freire et al. \cite{freire2021measuring} found that women are more likely to participate in organizing committees than as authors. The \textit{divinAI} project (2019) highlighted women's low representation in AI research \cite{gomez2024diversity}. Even when women engage in AI research, Stathoulopoulos and Mateos-Garcia \cite{stathoulopoulos2019gender} found that many tend to focus on societal issues—such as ethics, social impact, and policy—rather than purely technical domains.

Despite these insights, there remains a critical research gap in understanding the specific contributions of female authors in SSR. While diversity-related studies have explored broader trends and challenges, few have analyzed how women contribute to different roles of research within this field. Moreover, the impact of collaboration dynamics, particularly when research is led or supervised by female authors, has not been thoroughly investigated. Our study addresses this gap by providing detailed insights into the contributions of women in SSR and exploring how gender influences collaboration and leadership dynamics in this domain.

% ---------------------------------------------
\section{Methodology} \label{sec:methodology}
\begin{figure}[htb]
	\centering
	\includegraphics[width = 2.7in]{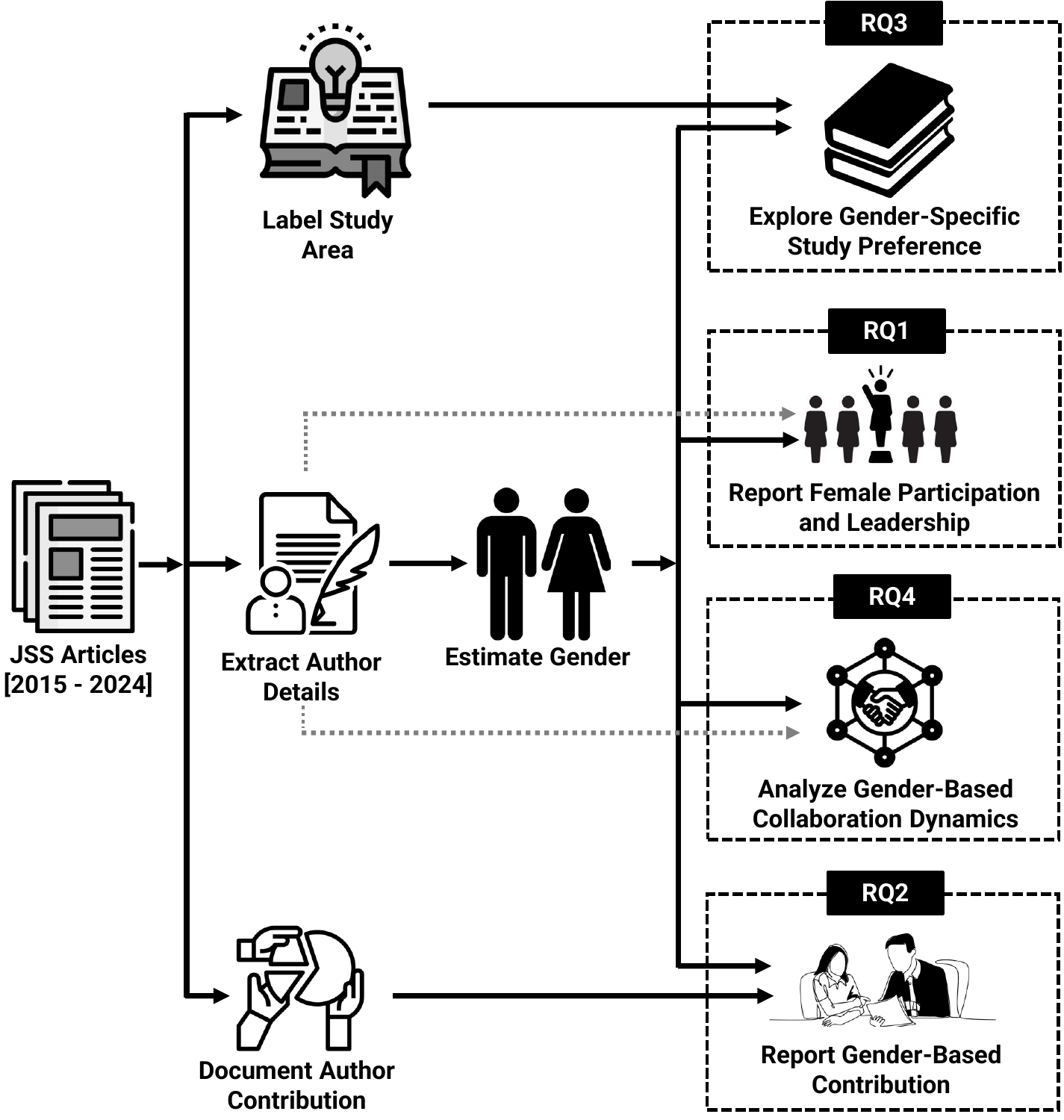}
	\caption{Study methodology}
	\label{fig:studyMethodology}
    \vspace{-4mm}
\end{figure}

% ---------------------------------------------
Fig.~\ref{fig:studyMethodology} shows the schematic diagram of our exploratory study. The individual steps are discussed as follows.

\begin{table*}[htb]
	\centering
	% \captionsetup{justification=centering, labelsep=newline}
	\caption{Details of Contributor Roles \cite{CRedit, allen2019can}}
	\label{table:contribution-details}
	\resizebox{6.8in}{!}{%
        \begin{tabular}{l|p{19cm}}
        \toprule

         \textbf{ID} & \textbf{Contributions} \\ \midrule
       
         C1 & \textbf{Conceptualization:} Refining the core ideas, establishing the research objectives, and outlining the overall direction or goals of the study.\\ \midrule
         
         C2 & \textbf{Data Curation:} Managing activities to produce metadata, scrubbing data, and maintaining research data for initial use and later re-use.\\ \midrule
       
         C3 & \textbf{Formal analysis:} Applying statistical, mathematical, computational, or other formal techniques to investigate study data. \\ \midrule
         
         C4 & \textbf{Funding acquisition:} Securing financial support for the project leading to this publication. \\ \midrule
        
         C5 & \textbf{Investigation:} Carrying out research and investigation, specifically performing experiments or collecting data/evidence.\\ \midrule
         
         C6 & \textbf{Methodology:} Developing or designing the methodology and creating models. \\ \midrule
       
         C7 & \textbf{Project administration:} Managing and coordinating the planning and execution of research activities. \\ \midrule

         C8 & \textbf{Resources:} Providing study materials, reagents, samples, patients, laboratory samples, animals, instruments, computing resources or other analytical tools. \\ \midrule
       
         C9 & \textbf{Software:} Programming and software development, including designing programs, implementing code and algorithms, and testing existing code components.\\ \midrule

        C10 & \textbf{Supervision:} Providing oversight and leadership for research planning and execution, including mentorship outside the core team.\\ \midrule

         C11 & \textbf{Validation} Verifying the replication and reproducibility of results, experiments, and other research outputs, either as part of the activity or as a separate process.\\ \midrule

         C12 & \textbf{Visualization:} Preparing, creating, and presenting the published work, with a focus on visualization and data presentation.\\ \midrule
     
         C13 & \textbf{Writing - original draft:} Preparing, creating, and presenting the published work, specifically by writing the initial draft, including any substantive translation.\\ \midrule

         C14 & \textbf{Writing - review \& editing:} Preparing, creating, and presenting the published work by original research group member, specifically through critical review, commentary, or revisions, including pre-and post-publication stages. \\ \bottomrule
        
        \end{tabular}
        }

\end{table*}

\subsection{Dataset Construction} \label{dataset_collection}
We collected 2,000 JSS articles from 2015 to 2024. JSS is one of the leading journals dedicated to software systems research, and it started in 1979.  
Another reason for choosing JSS is its inclusion of the \textit{CRedit authorship contribution statement} (\textit{CRedit} for short) section, which explicitly outlines authors' contributions—an important focus of our investigation (RQ2).
The Contributor Role Taxonomy (CRediT) was introduced to address the limitation of simple-ordered author lists, which fail to reflect the diverse contributions researchers make \cite{CRedit, allen2019can}. Thus, a 14-role taxonomy (refer to Table~\ref{table:contribution-details}) was developed to provide a more accurate representation and has been widely adopted since 2015 \cite{niso2022ansi}.
We then filtered the 2,000 articles based on the availability of the CRedit section. Table~\ref{tab:paper_stats} shows the breakdown of articles with and without the CRedit section. We found a total of 763 articles that had the CRedit section.  
Out of 763 articles, we excluded 364 that had only male authors and 15 that had only female authors. This left us with 384 articles featuring both female and male authors, which we selected for further analysis of contribution and collaboration patterns.

% \begin{table} [thb]\centering
%     \caption{Data Statistic}
%     \label{tab:paper_stats}
%     \resizebox{2in}{!}{
%     \begin{tabular}{c|c|c}
%         \toprule
%         \textbf{Year} & \begin{tabular}{@{}c@{}} \textbf{Total}\\ \textbf{Articles}\end{tabular} & \begin{tabular}{@{}c@{}} \textbf{\# of articles}\\ \textbf{with CRedit}\end{tabular} \\
%         \midrule
%         2015 & 181 & 0 \\ \midrule
%         2016 & 228 & 0 \\
%         \midrule
%         2017 & 207 & 0 \\ \midrule
%         2018 & 207 & 0 \\
%         \midrule
%         2019 & 174 & 0 \\ \midrule
%         2020 & 182 & 96\\
%         \midrule
%         2021 & 159 & 128 \\ \midrule
%         2022 & 182 & 116 \\
%         \midrule
%         2023 & 236 & 196 \\ \midrule
%         2024 & 244 & 227 \\
%         \toprule
%         \textbf{2015-2024} & \textbf{2000} & \textbf{763} \\
%         \bottomrule
%     \end{tabular}
%     }
% \end{table}

\begin{table*} [thb]\centering
    \caption{Dataset details}
    \label{tab:paper_stats}
    \resizebox{5in}{!}{%
    \begin{tabular}{r|c|c|c|c|c|c|c|c|c|c|c}
        \toprule
        \textbf{Year} & \textbf{2015} & \textbf{2016} & \textbf{2017} & \textbf{2018} & \textbf{2019} & \textbf{2020} & \textbf{2021} & \textbf{2022} & \textbf{2023} & \textbf{2024} & \textbf{2015-2024} \\
        \midrule
        \textbf{Total Articles} & 181 & 228 & 207 & 207 & 174 & 182 & 159 & 182 & 236 & 244 & \textbf{2000} \\
        \midrule
        \textbf{\# of articles with CRedit} & 0 & 0 & 0 & 0 & 0 & 96 & 128 & 116 & 196 & 227 & \textbf{763} \\
        \bottomrule
    \end{tabular}
    }
    \vspace{-3mm}
\end{table*}

\subsection{Analyzing the Prevalence of Female Authorship and Their Leadership}
\label{sec:Methodology-prevalence}
To analyze the prevalence of female authors, we first recorded the names of all authors for each article. We identified their genders using the biographies provided at the end of the articles. We then analyzed female authors' prevalence and their leadership and supervisory roles. Specifically:

$\bullet$ An article was labeled as female-led if the first author was female.

$\bullet$ An article was labeled as female-supervised if a female author held a supervisory role. 

\noindent The same criteria were applied to identify male-led and male-supervised studies. For analyzing collaboration dynamics (RQ\textsubscript{4}), we focused on studies with supervisors of a single gender. Out of 384 studies, we identified:

    $\bullet$ One hundred sixty-three articles with both male and female supervisors.
    
    $\bullet$ Two hundred twenty-one articles with a single-gender supervisor (45 female-supervised and 176 male-supervised).

\noindent To ensure clarity in studying gender-specific collaboration dynamics, we excluded the 163 articles with mixed-gender supervisors, narrowing the focus to the 221 single-gender-supervised studies.

\subsection{Analyzing the Contributions of Female Authors} 
\label{M2}
We analyzed the CRediT section of each article to identify the roles contributed by each author. The CRediT taxonomy includes 14 distinct roles (Table~\ref{table:contribution-details}) to which an author may contribute. For each role, we assigned a value of 1 if an author contributed and 0 otherwise.
We followed a paper-wise approach to calculate contribution ratios for female and male authors. For each paper, we determined the number of female and male contributors for a specific role and divided it by the total number of female or male authors in that paper. This paper-level analysis ensured fair representation by preventing the potential biases of an aggregated overall ratio, which might overestimate contributions in larger teams or underestimate them in smaller ones. By focusing on individual papers, we better understood the contribution dynamics within each study.

For a given paper:

    $\bullet$ Let $F_i$ be the number of female authors contributing to a specific role and $F_K$ the total number of female authors. The female contribution ratio ($CR_{female}$) for that role is: $CR_{female} = \frac{F_i}{F_K}$.
    
    $\bullet$ Similarly, let $M_i$ be the number of male authors contributing to a specific role, and $M_K$ be the total number of male authors. The male contribution ratio ($CR_{male}$) for that role is calculated as: $CR_{male} = \frac{M_i}{M_K}$.

\noindent For example, in a paper with two female authors and five male authors, if one female author and two male authors contributed to a specific role, the contribution ratios would be:

    $\bullet$ Female contribution ratio: $CR_{female} = \frac{F_1}{F_2} = \frac{1}{2} = 0.50$
    
    $\bullet$ Male contribution ratio: $CR_{male} = \frac{M_2}{M_5} = \frac{2}{5} = 0.40$

\noindent We used box plots to analyze and visualize the quartiles to understand the distribution of these contribution ratios, providing a clearer picture of the collaboration patterns across roles.

\subsection{Identifying Female Authors' Contributions in the Key Areas of Software Systems Research}

We conducted open coding on the selected 384 articles to identify the sub-domains of software systems research where female contributions and leadership are either prominent or underrepresented. The first two authors collaboratively labeled each article based on its title, abstract, and keywords, using open codes that reflected the study's focus. For instance, we coded the study by Zhu et al. \cite{zhu2021kill} with labels such as ``testing techniques'' and ``code quality and maintenance'' because these were the primary topics of the research. All 384 articles were manually labeled, and any disagreements during the coding process were resolved through discussion until a consensus was reached.

Consequently, a coding book was developed that included 107 labels. These labels were then organized into broader categories. For example, labels like ``AI Safety and Ethics,'' ``AI in Testing and Bug Detection,'' and ``AI in Development Tools'' were grouped under the main area of ``Applied AI in Software Engineering.'' This approach is similar to the one used by Sagdic et al. \cite{sagdic2024taxonomy}. As a result, we identified a total of 23 key areas within software systems.

After our initial round of coding, we identified that SSR in the JSS can be categorized into 13 key areas (i.e., scopes \cite{JSS_Scopes}), which closely resemble those in our codebook. To align with their framework, we first merged several of our key areas and subsequently renamed them. For instance, the JSS scope titled ``Methods and tools for software requirements, design, architecture, verification and validation, testing, maintenance, and evolution'' is quite broad. We combined our three main key areas--``Software Architecture and Design,'' ``Software Quality and Testing,'' and ``Software Maintenance and Evolution''—into this single category. This approach of integrating our identified key areas with JSS's established scopes ensures that our categorization is closely aligned with the journal's focus and accurately represents the recognized areas within SSR.

\begin{figure*}
    \centering
    \includegraphics[width=\textwidth]{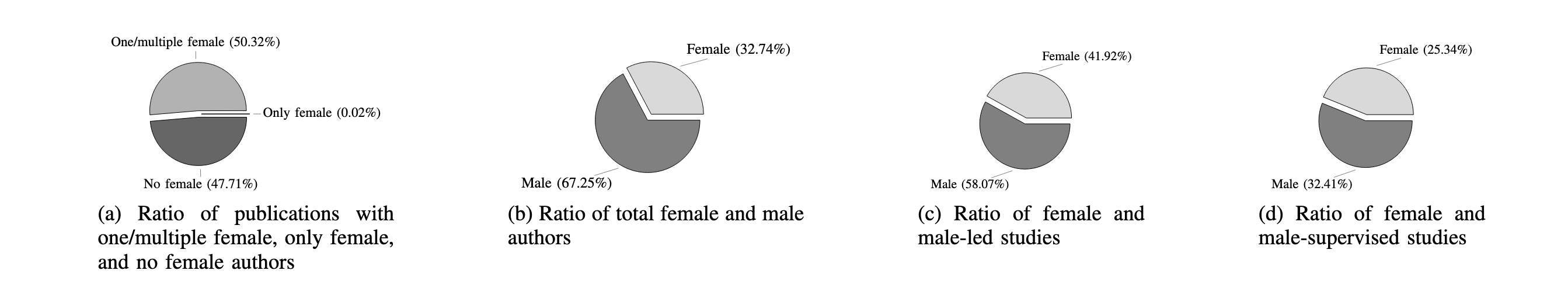}
    \caption{Comparison of different ratios across publications and author roles}
    \label{fig:side_by_side_pie_charts}
\end{figure*}

\subsection{Exploring the Collaboration Dynamics of Female Authors} \label{sec:methodology-collaboration}
We analyzed each author's affiliation, including their country and institution, to determine the type of collaboration involved in each study. Collaborations were categorized as follows:

    $\bullet$ \textbf{Local Collaboration:} All authors were from the same institution.
    
    $\bullet$ \textbf{National Collaboration:} Authors were from different institutions within the same country.
    
    $\bullet$ \textbf{International Collaboration:} Authors were from institutions in different countries.

We examined these collaboration patterns across all articles and then specifically focused on studies led or supervised by female authors, comparing their collaboration dynamics with those led or supervised by male authors. Additionally, we identified whether the collaborating institutions were academic or industrial to gain insights into the nature of these collaborations (e.g., industry-academia collaboration).

% -------------------------------------------------
\section{Study Findings} \label{sec:study-findings}
% -------------------------------------------------

\subsection{Participation and Leadership of Female Authors (RQ\textsubscript{1})} \label{sec:RQ1}

% Overall, about half of the articles have one or more female authors, whereas 47.71\% articles have no female authors at all, and only 15 out of 763 have only female authors (Fig~\ref{fig:authors_ratio}). 
% While a slight majority of articles feature female involvement, the fact that nearly half of the studies lack female participation points to a steady imbalance. 
Overall, approximately half of the articles include one or more female authors, while 47.71\% have no female authors at all. Furthermore, only 15 out of 763 articles are authored exclusively by females (Fig~\ref{fig:side_by_side_pie_charts}a). Although a slight majority of the articles involve female participation, the significant proportion of studies without any female authors highlights a persistent gender imbalance in the field.
% 
% To further examine gender representation, we analyzed 384 articles (as described in Section \ref{dataset_collection}) with both male and female authors. Fig.~\ref{fig:total_male_female} illustrates a low presence of females in SSR. Out of 1,771 total authors in 384 articles, only 580 authors are female, and 1,191 authors are male. This lower representation of women might extend into roles with greater influence, such as leadership and supervision.
To further investigate gender representation, we analyzed 384 articles that included both male and female authors. As shown in Fig.~\ref{fig:side_by_side_pie_charts}b, there is a noticeable underrepresentation of females in software systems research. Among the 1,771 authors across these articles, only 580 (32.7\%) are female, while 1,191 (67.3\%) are male. This disparity indicates that women may also be underrepresented in influential roles like leadership and supervision.

% We then analyzed the prevalence of women in leadership and supervisory roles. Fig.~\ref{fig:led_studies} shows that while female-led studies account for a significant portion (41.92\%), they are still outnumbered by male-led studies, which constitute 58.07\%. Similarly, Fig.~\ref{fig:supervised_studies} highlights that female authors are underrepresented in supervisory roles, with their participation nearly 8\% lower than that of male authors. This lower representation of women in leadership roles may be influenced by factors such as males perceiving that female leaders are less effective than their male counterparts \cite{roebuck2019organizational}. Such underrepresentation could hinder mentorship opportunities and career development for female researchers.

We then analyzed the prevalence of women in leadership and supervisory roles. As shown in Fig. \ref{fig:side_by_side_pie_charts}c, female-led studies represent an impressive percentage (41.92\%), but they remain fewer than male-led studies, which account for 58.07\%. Similarly, women are underrepresented in supervisory roles, with their participation nearly 8\% lower than that of men. This underrepresentation may be influenced by perceptions, such as the belief that female leaders are less effective than their male counterparts \cite{roebuck2019organizational}, potentially limiting mentorship opportunities and career growth for female researchers.

\smallskip
\noindent \textbf{Summary of RQ\textsubscript{1}:} More than 47\% of studies lack female authors, and the overall participation of female authors is low. Although over 40\% of studies have female authors in lead positions, the presence of female authors is 8\% less in supervisory positions than men.

% \tcbset{colback=gray!10!white,colframe=gray!50!black} 
% \begin{tcolorbox}[colframe=black!50, colback=white,left=0pt,right=1pt,top=1pt,bottom=1pt, arc=0.5pt] 
%     \textbf{Summary of RQ\textsubscript{1}:} More than 47\% of studies lack female authors, and the overall participation of female authors is low. Although over 40\% of studies have female authors in lead positions, the presence of female authors is 8\% less in supervisory positions than men.
% \end{tcolorbox}

\begin{figure}[htb]
	\centering
	\includegraphics[width = 3.4in]{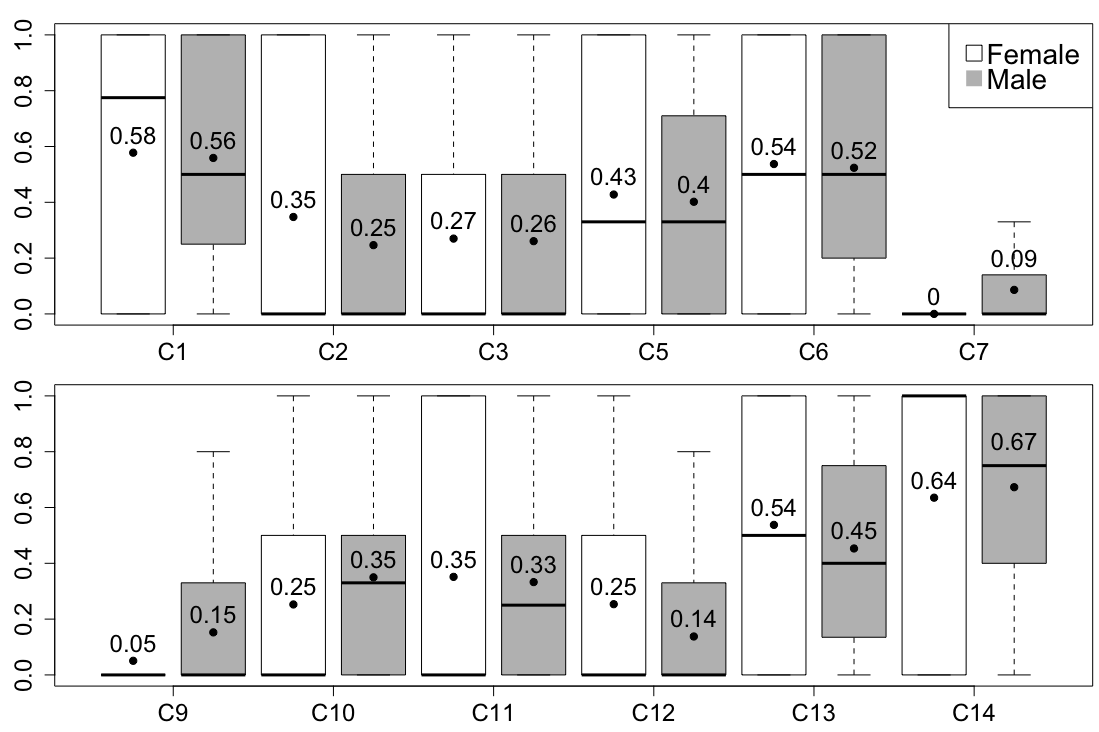}
	\caption{Contribution Ratio (\footnotesize{C1: Conceptualization; C2: Data Curation; C3: Formal analysis; C5: Investigation; C6: Methodology; C7: Project administration; C9: Software; C10: Supervision; C11: Validation; C12: Visualization; C13: Writing - original draft; C14: Writing - review \& editing}).}
	\label{fig:contribution ratio2}
    \vspace{-2mm}
\end{figure}

% ---------------------------------------------------------------------------------------------------------------
\subsection{Commitment and Contributions of Female Authors (RQ\textsubscript{2})} \label{sec:RQ2}
% ----------------------------------------------------------------------------------------------------------------

The previous section highlighted the overall prevalence of women's participation in SSR. This section analyzes their involvement across various contributory roles.
Fig.~\ref{fig:contribution ratio2} illustrates notable patterns in female authors' involvement in contribution roles. We excluded the funding acquisition and resources roles from the figure as participation from both male and female authors in these roles is minimal for quartile analysis. Specifically, 343 out of 384 studies had no female authors contributing to funding acquisition, and 312 out of 384 studies had no female authors involved in resource-related tasks. Similarly, 278 studies lacked male contributors in funding acquisition, and 297 studies had no male authors in the resources role. This suggests that both male and female authors contributed very little to funding and resource management tasks, with these roles mostly handled by very few individuals considered outliers (Section \ref{M2}).

For the rest of the contributory roles, our analysis reveals that female authors are contributing almost equally in formal analysis, investigation and methodology roles with males. However, their contributions are notable, particularly in conceptualization, writing the original draft, and reviewing it. For example, in at least half of the research articles, 78\% of female authors participated in conceptualization compared to male authors (50\%). Similarly, in the role of writing and reviewing the original draft, female authors are ahead of male authors. From 50\% to 100\% of female authors and from 40\% to 75\% of male authors are contributing in at least half of the articles. This finding can also be supported by a study based on top economics journals by Hengel \cite{hengel2017publishing}, which found that female economists often excel in writing clarity, surpassing men in that area.

On the contrary, female participation sharply declines in supervision and validation roles.  In at least half of the articles, female authors did not contribute to either of these roles at all, while male authors contributed in supervision and validation roles at rates of 33\% and 25\%, respectively. Lindahl et al. \cite{lindahl2021importance} found that men often produce more research output than women during and after doctoral studies. Since women are less often involved in supervision and validation, they may have fewer chances to co-author with senior researchers and build strong networks.
However, female authors also contribute in a relatively balanced way with male authors in data curation, project administration, software, and visualization roles, with average contributions ranging from 0.05 to 0.35 compared to male authors’ range of 0.09 to 0.35.

\smallskip
\noindent \textbf{Summary of RQ\textsubscript{2}:}
    The results indicate that female authors contribute almost equally in formal analysis, investigation, and methodology, excelling in conceptualization, writing, and reviewing. However, their contributions are less recognized in supervision and validation.

% \begin{tcolorbox}[colframe=black!50, colback=white,left=0pt,right=1pt,top=1pt,bottom=1pt, arc=0.5pt]
%     \textbf{Summary of RQ\textsubscript{2}:}
%     The results indicate that female authors contribute almost equally in formal analysis, investigation, and methodology, excelling in conceptualization, writing, and reviewing. However, their contributions are less recognized in supervision and validation. 
%     % Additionally, female authors show a relatively balanced involvement with male authors in data curation, project administration, software, and visualization roles.
% \end{tcolorbox}

% ---------------------------------------------------------------------------------------------------------------
\subsection{Leadership and Contribution of Female Authors' Across Key Areas (RQ\textsubscript{3})} \label{sec:RQ3}
% ----------------------------------------------------------------------------------------------------------------

Previously, we found women contribute nearly as much as men. In this section, we analyze the broader context to explore the research domains of these articles, focusing on where female researchers are more or less involved than their male counterparts.
A quantitative comparison in Table~\ref{tab:dataset_summary} shows that men-led studies outnumber women in the key areas of Artificial Intelligence (AI), data analytics, and big data in software systems research (\textit{S5}). Specifically, $22.4\%$ of men-led studies were in \textit{S5}, while only $12.4\%$ of women-led studies fell into this category.
% Furthermore, our Chi-Square \cite{mchugh2013chi} analysis also yielded a statistically significant disparity ($p=0.018$) specifically within the subdomain of \textit{S5}. 
Interestingly, no other major areas in SSR exhibited such a gender-based difference. This trend is concerning; first-authored studies by women in AI for software systems research are significantly lower than those by men, highlighting the scarcity of female leadership in the most progressive sector of software systems research.

 Conversely, we found that women are leading the majority of research in the key area of human factors and management in software development (\textit{S4}), with $33.3\%$ under female supervision compared to just $10.8\%$ for male supervision. Furthermore, we have identified a statistically substantial difference ($p=0.0005$), specifically in the \textit{S4} subdomain. Once again, no other major areas in software systems exhibit such a significant gender-based difference in supervision. Cavero et al. \cite{catolino2019gender} also found that women are indeed more engaged in human-centered areas of computer science, and this pattern seems to extend into research on software systems. This trend may stem from historical perceptions that women are more skilled in ``people-oriented'' roles \cite{eagly2013female, korabik1989should}, which may have influenced their research choices toward human-centric studies rather than technical fields like AI.

\smallskip
\noindent \textbf{Summary of RQ\textsubscript{3}:}  The results show that female first authors are less common than males in AI in software systems research. However, female faculty supervise research on the human aspects of software more, indicating a focus on social implications, while men are more inclined to emphasize improving systems with AI.

% \begin{tcolorbox}[colframe=black!50,colback=white,left=0pt,right=1pt,top=1pt,bottom=1pt, arc=0.5pt]
%     \textbf{Summary of RQ\textsubscript{3}:}  The results show that female first authors are less common than males in AI in SE. However, female faculty supervise research on the human aspects of software more, indicating a focus on social implications, while men are more inclined to emphasize improving systems with AI.
% \end{tcolorbox}

% Study Area, Female Led %, Male Led %, Female Led difference compared to Male Led, Female Supervised %, Male Supervised %, Female Supervised difference compared to Male supervised, Total difference

\begin{table*}[t] 
    \centering
    \caption{Gender-Based Leadership and Supervision of Articles in Key Areas of Software Systems Research}
    \resizebox{6.5in}{!}{
    \begin{tabular}{l|l|r|r|r|r}
        \toprule
        \begin{tabular}{@{}c@{}} ID\end{tabular} & 
        \begin{tabular}{@{}c@{}} Key Areas\end{tabular} & 
        \begin{tabular}{@{}c@{}} Female Led \\(\textit{out of 161 papers}) \end{tabular} &
        \begin{tabular}{@{}c@{}} Male Led \\(\textit{out of 223 papers})\end{tabular} &
        \begin{tabular}{@{}c@{}} Female Sup \\(\textit{out of 45 papers})\end{tabular} & 
        \begin{tabular}{@{}c@{}} Male Sup \\(\textit{out of 176 papers})\end{tabular} \\
        \midrule
         S1 &\begin{tabular}{@{}p{10cm}@{}} Methods and tools for software requirements, design, architecture, verification and validation, testing, maintenance and evolution\end{tabular}
         & 38 (23.6\%) & 64 (28.7\%) & 9 (20.0\%) & 46 (26.1\%)  \\ \midrule
   
         S2 &\begin{tabular}{@{}p{10cm}@{}} Agile, model-driven, service-oriented, open source and global software development\end{tabular} & 8 (5.0\%) &10 (4.5\%)  & 0 (0\%) & 13 (7.4\%) \\  
        \midrule
         S3 & \begin{tabular}{@{}p{10cm}@{}} Approaches for cloud/fog/edge computing and virtualized systems \end{tabular}  & 11 (6.8\%) & 14 (6.3\%) &2 (4.4\%) &9 (5.1\%) \\ \midrule

         S4 &\begin{tabular}{@{}p{10cm}@{}} Human factors and management concerns of software development \end{tabular}& 28 (17.4\%) &29  (13\%) &15 (33.3\%) &19 (10.8\%)  \\
        \midrule
         S5 &\begin{tabular}{@{}p{10cm}@{}} Artificial Intelligence, data analytics and big data applied in software engineering \end{tabular}&20 (12.4\%) &50 (22.4\%) &7 (15.6\%) &42 (23.9\%) \\  \midrule
         
         S6 &\begin{tabular}{@{}p{10cm}@{}} Metrics and evaluation of software development resources \end{tabular} &0 (0\%) &2 (0.9\%) &0 (0\%) & 0 (0\%) \\ 
         
         \midrule
         S7 & \begin{tabular}{@{}p{10cm}@{}} DevOps, continuous integration, build and test automation \end{tabular} &5 (3.1\%) &1 (0.4\%) &1 (2.2\%) &3 (1.7\%) \\ \midrule
         S8 & \begin{tabular}{@{}p{10cm}@{}} Business and economic aspects of software development processes \end{tabular}&3 (1.9\%) &3 (1.3\%) & 0  (0\%)&1 (0.6\%) \\  \midrule
         S9 & \begin{tabular}{@{}p{10cm}@{}} Software Engineering education\end{tabular} &7 (4.3\%) &6 (2.7\%) &3 (6.7\%) &2 (1.1\%) \\ \midrule
         S10 & \begin{tabular}{@{}p{10cm}@{}} Ethical/societal aspects of Software Engineering \end{tabular}  &3 (1.9\%)  &1 (0.4\%) & 0 (0\%) &1 (0.6\%)\\
        \midrule
         S11 & \begin{tabular}{@{}p{10cm}@{}} Software Engineering for AI systems \end{tabular} &3 (1.9\%) &2 (0.9\%) &2 (4.4\%) &3 (1.7\%) \\ \midrule
         S12 & \begin{tabular}{@{}p{10cm}@{}} Software Engineering for Sustainability\end{tabular} &2 (1.2\%) &0 (0\%) & 0 (0\%)& 0 (0\%) \\ 
         \midrule
         S13 & \begin{tabular}{@{}p{10cm}@{}} Methods and tools for empirical software engineering research \end{tabular} &33 (20.5\%) &41 (18.4\%) &6 (13.3\%) & 21.0 (37\%) \\ \bottomrule
         
         % \cmidrule(lr){3-6}
         %    & \begin{tabular}{@{}p{10cm}@{}} \raggedleft \textbf{Total Articles} \end{tabular} & 161 & 223 & 45 & 176 \\ \cmidrule(lr){3-6}
  \end{tabular}
  }
    \label{tab:dataset_summary}
\end{table*}

% ---------------------------------------------------------------------------------------------------------------
\subsection{Collaboration Dynamics of Female Authors (RQ\textsubscript{4})} \label{sec:RQ4}
% ----------------------------------------------------------------------------------------------------------------
Previously we found that women are underrepresented in the leading and supervisory roles. Moreover, despite their underrepresentation, they have significant involvement in the contributory roles. Therefore, we further attempted to explore how female authors can influence different collaboration patterns, especially when they are in the lead or supervisory position.   
At first, we explored the overall collaboration pattern for all 384 studies. We found that over 54.43\% (209 out of 384) of the articles have international collaboration. However, local (93 out of 384 studies) and national (82 out of 384 studies) collaboration have comparatively balanced ratios ranging from 21\% to 24\%.  

% \begin{figure}[!htb]
% 	\centering
% 	\resizebox{2in}{!}{
%     \begin{tikzpicture}
%     %   \pie [polar, explode=0.1, color={black!5, black!15, black!25}, text=legend]
%     \pie[explode=0.2, text=pin, sum = auto, color={black!10, black!30, black!50}, number in legend] %hide number
%         { 24.22/\Huge{Local (24.22\%)},
%           21.35/\Huge{National (21.35\%)},
%           54.43/\Huge{International (54.43\%)}
%           }
%     \end{tikzpicture}
%     }
% 	\caption{Collaboration dynamics of all the studies}
% 	\label{fig:collaboration_all_studies}
% % 	\vspace{-3mm}
% \end{figure}

Next, we investigated the collaboration dynamics when a female author holds the lead or supervisory position and conducted a similar analysis for male authors as well. For investigating the collaboration dynamic in the supervisory positions, we focused on studies where either a female or a male was in a supervisory position (Section, \ref{sec:Methodology-prevalence}).

\begin{figure}[!htb]
\centering
    \pgfplotstableread{
    1	21.73	24.85	53.42
    2   26.01   18.83   55.16
    3   33.33   13.33   53.33
    4   25.00   24.43   50.57
    }\datatable
    \subfloat{
    \resizebox{2.8in}{!}{%
        \begin{tikzpicture}
        \begin{axis}[
            xtick=data,
            xticklabels={F-led,M-led,F-supervised,M-supervised},
            enlarge y limits=false,
            enlarge x limits=0.15,
            % nodes near coords,
            ymin=0,ymax=105,
            ybar,
            bar width=0.4cm,
            width=3.6in,
            height = 2in,
            ytick={0,20,...,100},
            yticklabels={0\%,20\%,40\%,60\%,80\%,100\%,},
            ymajorgrids=false,
            %	xminorgrids=true,
            % 	yticklabel style={font=\small},
            % 	xticklabel style={font=\footnotesize, /pgf/number format/fixed},	
            major x tick style = {opacity=0},
            minor x tick num = 1,    
            minor tick length=1ex,
            legend style={
                    at={(0.65,1)},
            	font=\footnotesize,
                    % 	legend pos = outer north east,
             	% legend pos=north east,
                    %   anchor=west,
                    % 	cells={align=left},
            	legend cell align=left,
                    anchor=north,legend columns=-1,
                    draw=none
                },
            nodes near coords style={rotate=90,  anchor=west}, font=\small,
            nodes near coords =\pgfmathprintnumber{\pgfplotspointmeta}\%
            %nodes near coords*={\pgfmathprintnumber[precision=2]\pgfplotspointmeta \%}
            ]
            \addplot[draw=black!80, fill=black!10] table[x index=0,y index=1] \datatable;
            \addplot[draw=black!80, fill=black!50] table[x index=0,y index=2] \datatable;
            \addplot[draw=black!80, fill=black!90] table[x index=0,y index=3] \datatable;

            \legend	{Local, National, International}
        
        \end{axis}
        \end{tikzpicture}
        }
    }
\caption{Collaboration dynamics across authors' roles}
\label{fig:pie_charts}
\end{figure}
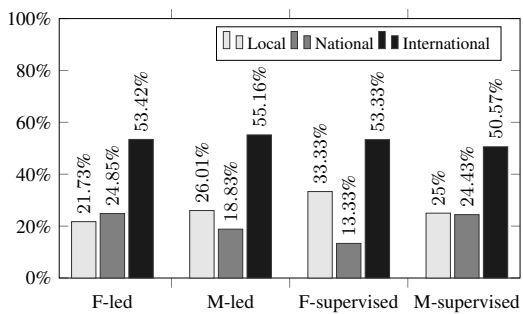

Fig.~\ref{fig:pie_charts} shows that both female-led and male-led studies prioritize international collaboration ranging from 53\% to 55\%. 
However, slightly notable differences exist in the local and national collaboration. Female-led studies show 6\% more national collaboration but 5\% less local collaboration compared to male-led studies. 
Similarly, we also found that both female and male-supervised studies heavily emphasize international collaborations.
However, unlike female-led studies, female-supervised studies have a notably higher ratio of local collaboration (33.33\%), with a 20\% difference over national collaboration (13.33\%). In contrast, male-supervised studies maintain a more balanced distribution between local and national collaborations (up to 25\%).
The higher difference in local collaboration for female-supervised studies may be attributed to the tendency of female supervisors to prioritize local collaborations, likely due to strong connections within their immediate academic or professional communities. In contrast, female lead authors seem to favor national collaborations slightly more, possibly because they are relatively new to the profession and seek to connect with researchers in similar fields to expand their networks. However, female supervisors seem less willing to conduct national collaborations, potentially because their research domains may not align with those of other lab supervisors at the national level.

Lastly, we examined collaboration types at the industrial and academic levels. Both female-led and female-supervised studies demonstrate similar levels of industrial collaboration, ranging from 25\% to 27\%. Interestingly, male-led and male-supervised studies follow a similar trend, showing a consistent engagement in industry collaboration. This similarity across both female and male leadership types highlights a consistent preference for industrial partnership, regardless of gender.

\smallskip
\noindent \textbf{Summary of RQ\textsubscript{4}:} Both female-led and female-supervised studies prioritized international collaborations. Male-led and male-supervised studies had a balanced mix of local and national collaborations, while female-supervised studies showed 20\% more local than national collaboration. No significant differences were found in industrial versus academic collaborations across all study types.

% \begin{tcolorbox}[colframe=black!50, colback=white,left=0pt,right=1pt,top=1pt,bottom=1pt, arc=0.5pt]
%     \textbf{Summary of RQ\textsubscript{4}:} Both female-led and female-supervised studies prioritized international collaborations. Male-led and male-supervised studies had a balanced mix of local and national collaborations, while female-supervised studies showed 20\% more local than national collaboration. No significant differences were found in industrial versus academic collaborations across all study types.
% \end{tcolorbox}

% -----------------------
\section{Key findings}
% -----------------------
% In this section, we discuss the key findings of this study.

\begin{figure}[pt]
\centering
  \resizebox{1.4in}{!}{%
    \begin{tikzpicture}
        \begin{axis}[
            width=2in,
            height=2in,
            axis lines=left,
            grid=both,
            legend cell align=left,
            legend style={
            	at={(0.65,0.98)},
            	font=\small,
                draw= none
            },
            ticklabel style={font=\small},
            xmin=1, 
            ymax=70,
            ymin=30,
            xmax=5,
            xtick={1,2,3,4,5},
            xticklabels={
                        2020,
                        2021,
                        2022,
                        2023,
                        2024,
                        },
            x tick label style={rotate=90,anchor=east},
            ytick={20, 30,...,70},
            yticklabels={25\%,30\%,35\%,40\%,45\%,50\%,55\%,60\%,65\%,70\%},
            legend entries={
                female,
                male
            },
            cycle list name = auto % color,exotic,black white,mark list,mark list*,linestyles,linestyles*,auto
        ]
    \addplot[color=black!40!red, mark=*, thick, mark options={scale=1.0}, solid]
    coordinates {
                (1,35.71)
                (2,53.33)
                (3,43.75)
                (4,43.30)
                (5,33.96)
                };
    \addplot[color=black!40!blue, mark=square*, thick, mark options={scale=1.0}, solid]
    coordinates {
                (1,64.29)
                (2,46.67)
                (3,56.25)
                (4,56.70)
                (5,66.04)
                };
    \end{axis}
    \end{tikzpicture}
    }
\caption{Female-led and male-led studies over the last 5 years}
\vspace{-3mm}
\label{fig:total_female_led_male_led_studies}
\end{figure}
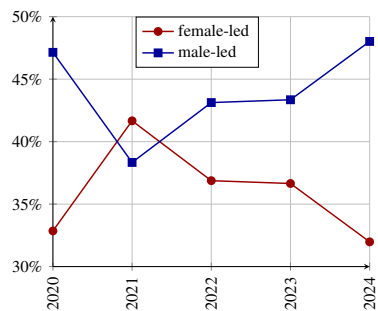

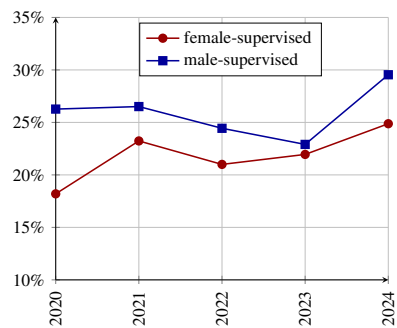
\begin{figure}[pt]
\centering
  \resizebox{1.4in}{!}{%
    \begin{tikzpicture}
    \begin{axis}[
        width=2in,
        height=2in,
        axis lines=left,
        grid=both,
        legend cell align=left,
        legend style={
        	at={(0.87,0.98)},
        	font=\small,
            draw = none
        },
        ticklabel style={font=\small},
        xmin=1, 
        ymax=40,
        ymin=15,
        xmax=5,
        xtick={1,2,3,4,5},
        xticklabels={
                    2020,
                    2021,
                    2022,
                    2023,
                    2024
                    },
        x tick label style={rotate=90,anchor=east},
        ytick={15,20,25,30,35, 40},
        yticklabels={15\%,20\%,25\%,30\%,35\%, 40\%},
        legend entries={
            female,
            male
        },
        cycle list name = auto % color,exotic,black white,mark list,mark list*,linestyles,linestyles*,auto
    ]
    
    \addplot[color=black!40!red, mark=*, thick, mark options={scale=1.0}, solid]
    coordinates {
                (1,16.39)
                (2,26.47)
                (3,22.00)
                (4,23.90)
                (5,29.75)
                };
    \addplot[color=black!40!blue, mark=square*, thick, mark options={scale=1.0}, solid]
    coordinates {
                (1,32.54)
                (2,33.02)
                (3,28.88)
                (4,25.81)
                (5,39.08)
                };
    \end{axis}
    \end{tikzpicture}
    }
\caption{Female-supervised and male-supervised studies over the last 5 years}
\vspace{-3mm}
\label{fig:total_female_supervised_male_supervised_studies}
\end{figure}

\noindent\textbf{Fading Women, Rising Disparity.} Our studies found that over 47\% (364 out of 763) of studies do not have any female authors at all (Section~\ref{sec:RQ1}). Even in those studies, female participants are significantly low. While female-led studies have a higher ratio (41.92\%) than female-supervised studies (25.34\%), Fig.~\ref{fig:total_female_led_male_led_studies} shows that female-led studies have been declining over the past five years. After reaching its highest point at 53.33\% in 2021, the percentage of female-led studies has steadily declined through 2024. Conversely, female-supervised studies (Fig.~\ref{fig:total_female_supervised_male_supervised_studies}) have gradually increased, but their ratio is still lower than male-supervised studies. Even when male-supervised studies were at their lowest in 2023, female-supervised studies did not reach a similarly high level. 
These findings align with research by Kohl et al. \cite{kohl2021challenges}, which highlights the challenges women face in software engineering management roles. Women in leadership roles often feel they must work harder than their male peers to gain the same level of recognition. There is a perception that when men perform the same work, it is praised, whereas when women do it, it is viewed with question, highlighting the additional pressures women face in these roles.

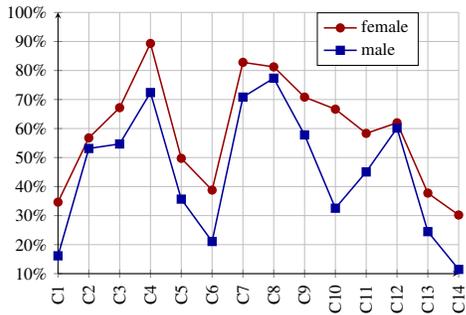
\begin{figure}[pt]
\centering
  \resizebox{2.5in}{!}{%
    \begin{tikzpicture}
        \begin{axis}[
            width=3.5in,
            height=2.5in,
            axis lines=left,
            grid=both,
            legend cell align=left,
            legend style={
            	at={(0.89,0.98)},
            	font=\small,
                draw= none
            },
            ticklabel style={font=\small},
            xmin=1, 
            ymax=100,
            ymin=10,
            xmax=14,
            xtick={1,2,3,4,5,6,7,8,9,10,11,12,13,14},
            xticklabels={
                        C1,
                        C2,
                        C3,
                        C4,
                        C5,
                        C6,
                        C7,
                        C8,
                        C9,
                        C10,
                        C11,
                        C12,
                        C13,
                        C14,
                        },
            x tick label style={rotate=90,anchor=east},
            ytick={10,20,...,100},
            yticklabels={10\%,20\%,30\%,40\%,50\%,60\%,70\%,80\%,90\%,100\%},
            legend entries={
                female,
                male
            },
            cycle list name = auto % color,exotic,black white,mark list,mark list*,linestyles,linestyles*,auto
        ]
    \addplot[color=black!40!red, mark=*, thick, mark options={scale=1.0}, solid]
    coordinates {
                (1,34.64)
                (2,56.77)
                (3,67.19)
                (4,89.32)
                (5,49.74)
                (6,38.80)
                (7,82.81)
                (8,81.25)
                (9,70.83)
                (10,66.67)
                (11,58.33)
                (12,61.98)
                (13,37.76)
                (14,30.21)};
            
    \addplot[color=black!40!blue, mark=square*, thick, mark options={scale=1.0}, solid]
    coordinates {
                (1,16.15)
                (2,53.13)
                (3,54.69)
                (4,72.40)
                (5,35.68)
                (6,21.09)
                (7,70.83)
                (8,77.34)
                (9,57.81)
                (10,32.55)
                (11,45.05)
                (12,60.16)
                (13,24.48)
                (14,11.46)
                };
    \end{axis}
    \end{tikzpicture}
    }
\caption{Ratio of studies where females are not involved in the specific contributions}
\vspace{-3mm}
\label{fig:no_contribution}
\end{figure}

\noindent\textbf{Women in Software Systems Research: Still Undervalued.} Our study found that out of 14 contributory roles, women are involved equally or similarly in most of the roles (Section \ref{sec:RQ2}). Moreover, their contribution is significantly higher in outlining the study objective and preparing the manuscript. However, as shown in Fig.~\ref{fig:no_contribution},  the number of studies where women had no contribution to a specific role is consistently lower than that of men. The lower rate of contribution by women in SSR teams may stem from their undervaluation within these groups. Research indicates that women often face challenges such as gender bias, lack of recognition, and limited access to leadership opportunities, which can hinder their active participation and contribution. For instance, a study by Guzm{\'a}n \cite{guzman2024mind} highlights that women in the software industry frequently encounter socio-cultural challenges that discourage their involvement, leading to feelings of isolation and undervaluation.

% In 2011, Marc Andreessen, co-author of Mosaic the first widely used web-browser famously said ``Software is eating the world'' pointing out the companies that embraced software in 2011 will be the current market leaders. Fast forward now, the top 5 market capitalization companies worldwide all offered some type of software solutions. Now, ``GenAI is eating software'' and impacted all the phases of SE including development, testing and maintenance. Therefore, in next decade the AI startups might lead the lead. MOreover, the implications of AI in SE could be momentous, with predictions indicating a surge in productivity ranging from 20\% to 45\% \cite{mckinseyEconomicPotential}.  Therefore, in next decade the AI startups might lead the lead. Similarly at the individual level, our research statistically shows that studies on AI in SE is a male dominant research domain in software systems. Therefore, female researcher might in a trajectory to fall behind even more in near future where most SSR are predicted to have some AI in it. 
\noindent\textbf{Unequal footing of females in AI for software systems research.}
In 2011, Marc Andreessen, co-author of the first widely used web browser, stated, ``Software is eating the world,'' \cite{andreessen2011software} noting that companies embracing software at that time would become market leaders. Today, the top five companies by market capitalization all offer software solutions \cite{foolLargestCompanies}. 
Presently, Gen AI is transforming software across all engineering phases—development, testing, and maintenance. Thus, companies embracing AI may be the future leaders and may boost developer productivity by 20\% to 45\% \cite{mckinseyEconomicPotential}, adding significant value to their AI tools. Companies that do not adopt AI into their workflows might be wiped out.
Similarly, at the individual level, our research statistically shows that studies on AI in software systems research are male-dominant. As a result, female researchers may risk falling further behind, especially considering that future software systems are expected to increasingly incorporate AI.
\smallskip
% -------------------------------------------------------
\section{Threats to Validity} \label{threats_to_validity}
% -------------------------------------------------------

Threats to \textbf{external validity} relate to the generalizability of our findings.
Our study focuses on software systems research using the JSS as a key source. While other journals cover related topics, they often lack a dedicated section highlighting author contributions. By analyzing all JSS articles since the introduction of CRediT, we ensured thorough coverage for this journal. Our findings can be applied to other journals using the CRediT taxonomy but may not generalize to fields with different contribution recording methods.

Threats to \textbf{internal validity} relate to experimental errors and biases.
Errors in gender prediction and biases in manual categorization of research areas pose risks to internal validity. To reduce bias, the first two authors worked together to code articles, grouped codes into themes, and resolved disagreements through discussion. For consistency, we also aligned our themes with established JSS research areas.

Another issue is our binary approach to gender identification, as biographies only mention male and female genders. Including other gender identities would require direct input from authors, which was beyond the scope of this study. We recommend that future research adopt inclusive methods to account for non-binary and other genders for a more accurate analysis.

\section{Conclusion} \label{conclusion}
% -------------------------------------
This study investigates women's participation, leadership, contributions, and collaboration dynamics in SSR. We analyzed 384 JSS articles using the CRedit taxonomy and discovered that nearly 48\% of the studies had no female authors. This finding highlights a continuing underrepresentation of women in the field. Moreover, female-led studies accounted for 41.92\% of the total but have declined over the past five years. Meanwhile, female-supervised studies have increased gradually, though they still trail behind male-supervised studies.
Women have made significant contributions in roles like conceptualization, writing, and reviewing, but their participation in supervisory and validation roles is still low. Female-led studies are less common in AI-focused research, while studies supervised by women are more frequent in human-centric fields.
Collaboration patterns differed between female-supervised and female-led studies. Female-supervised studies favored local collaborations, whereas female-led studies preferred national collaborations. These findings reveal biases affecting women in different roles and highlight the need for targeted efforts to promote gender equity in SSR.

\section{Acknowledgement}
This research is supported in part by the Natural Sciences and Engineering Research Council of Canada (NSERC) Discovery Grants program, the Canada Foundation for Innovation's John R. Evans Leaders Fund (CFI-JELF), and by the industry-stream NSERC CREATE in Software Analytics Research (SOAR).

\bibliographystyle{IEEEtran}
{
    \footnotesize
    \bibliography{GE_ICSE}

\begin{thebibliography}{10}
\providecommand{\url}[1]{#1}
\csname url@samestyle\endcsname
\providecommand{\newblock}{\relax}
\providecommand{\bibinfo}[2]{#2}
\providecommand{\BIBentrySTDinterwordspacing}{\spaceskip=0pt\relax}
\providecommand{\BIBentryALTinterwordstretchfactor}{4}
\providecommand{\BIBentryALTinterwordspacing}{\spaceskip=\fontdimen2\font plus
\BIBentryALTinterwordstretchfactor\fontdimen3\font minus \fontdimen4\font\relax}
\providecommand{\BIBforeignlanguage}[2]{{%
\expandafter\ifx\csname l@#1\endcsname\relax
\typeout{** WARNING: IEEEtran.bst: No hyphenation pattern has been}%
\typeout{** loaded for the language `#1'. Using the pattern for}%
\typeout{** the default language instead.}%
\else
\language=\csname l@#1\endcsname
\fi
#2}}
\providecommand{\BIBdecl}{\relax}
\BIBdecl

\bibitem{albusays2021diversity}
K.~Albusays, P.~Bjorn, L.~Dabbish, D.~Ford, E.~Murphy-Hill, A.~Serebrenik, and M.-A. Storey, ``The diversity crisis in software development,'' \emph{IEEE Software}, vol.~38, no.~2, pp. 19--25, 2021.

\bibitem{blincoe2019perceptions}
K.~Blincoe, O.~Springer, and M.~R. Wrobel, ``Perceptions of gender diversity's impact on mood in software development teams,'' \emph{Ieee Software}, vol.~36, no.~5, pp. 51--56, 2019.

\bibitem{happe2021frustrations}
L.~Happe and B.~Buhnova, ``Frustrations steering women away from software engineering,'' \emph{IEEE Software}, vol.~39, no.~4, pp. 63--69, 2021.

\bibitem{vasilescu2015gender}
B.~Vasilescu, D.~Posnett, B.~Ray, M.~G. van~den Brand, A.~Serebrenik, P.~Devanbu, and V.~Filkov, ``Gender and tenure diversity in github teams,'' in \emph{Proceedings of the 33rd annual ACM conference on human factors in computing systems}, 2015, pp. 3789--3798.

\bibitem{earley2000creating}
C.~P. Earley and E.~Mosakowski, ``Creating hybrid team cultures: An empirical test of transnational team functioning,'' \emph{Academy of Management journal}, vol.~43, no.~1, pp. 26--49, 2000.

\bibitem{jehn1999differences}
K.~A. Jehn, G.~B. Northcraft, and M.~A. Neale, ``Why differences make a difference: A field study of diversity, conflict and performance in workgroups,'' \emph{Administrative science quarterly}, vol.~44, no.~4, pp. 741--763, 1999.

\bibitem{hyde2014gender}
J.~S. Hyde, ``Gender similarities and differences,'' \emph{Annual review of psychology}, vol.~65, no.~1, pp. 373--398, 2014.

\bibitem{bastarrica2018affirmative}
M.~C. Bastarrica, N.~Hitschfeld, M.~M. Samary, and J.~Simmonds, ``Affirmative action for attracting women to stem in chile,'' in \emph{Proceedings of the 1st International Workshop on Gender Equality in Software Engineering}, 2018, pp. 45--48.

\bibitem{bello2021smart}
A.~Bello, T.~Blowers, S.~Schneegans, and T.~Straza, ``To be smart, the digital revolution will need to be inclusive,'' \emph{Excerpt from the UNESCO scient report}, 2021.

\bibitem{felizardo2021global}
K.~R. Felizardo, A.~M. Ramos, C.~d. O~Melo, {\'E}.~F. de~Souza, N.~L. Vijaykumar, E.~Y. Nakagawa \emph{et~al.}, ``Global and latin american female participation in evidence-based software engineering: a systematic mapping study,'' \emph{Journal of the Brazilian Computer Society}, vol.~27, no.~1, pp. 1--22, 2021.

\bibitem{spangenberg2021holistic}
L.~Spangenberg and H.~W. Pretorius, ``Holistic factors that impact the under-representation of women in ict: A systematic literature review,'' in \emph{Proceedings of Fifth International Congress on Information and Communication Technology: ICICT 2020, London, Volume 1}.\hskip 1em plus 0.5em minus 0.4em\relax Springer, 2021, pp. 195--209.

\bibitem{cundiff2016gender}
J.~L. Cundiff and T.~K. Vescio, ``Gender stereotypes influence how people explain gender disparities in the workplace,'' \emph{Sex Roles}, vol.~75, no.~3, pp. 126--138, 2016.

\bibitem{hyrynsalmi2019underrepresentation}
S.~M. Hyrynsalmi, ``The underrepresentation of women in the software industry: thoughts from career-changing women,'' in \emph{2019 IEEE/ACM 2nd International Workshop on Gender Equality in Software Engineering (GE)}.\hskip 1em plus 0.5em minus 0.4em\relax IEEE, 2019, pp. 1--4.

\bibitem{trinkenreich2022women}
B.~Trinkenreich, I.~Wiese, A.~Sarma, M.~Gerosa, and I.~Steinmacher, ``Women’s participation in open source software: A survey of the literature,'' \emph{ACM Transactions on Software Engineering and Methodology (TOSEM)}, vol.~31, no.~4, pp. 1--37, 2022.

\bibitem{kazmi2014women}
A.~Kazmi, ``Women managers in different types of organisations: A representative research review,'' \emph{Journal of Entrepreneurship and Management}, vol.~3, no.~1, 2014.

\bibitem{maheshwari2023review}
G.~Maheshwari, ``A review of literature on women’s leadership in higher education in developed countries and in vietnam: Barriers and enablers,'' \emph{Educational Management Administration \& Leadership}, vol.~51, no.~5, pp. 1067--1086, 2023.

\bibitem{cavero2015evolution}
J.~M. Cavero, B.~Vela, P.~C{\'a}ceres, C.~Cuesta, and A.~Sierra-Alonso, ``The evolution of female authorship in computing research,'' \emph{Scientometrics}, vol. 103, pp. 85--100, 2015.

\bibitem{boekhout2021gender}
H.~Boekhout, I.~van~der Weijden, and L.~Waltman, ``Gender differences in scientific careers: A large-scale bibliometric analysis,'' \emph{arXiv preprint arXiv:2106.12624}, 2021.

\bibitem{bano2019gender}
M.~Bano and D.~Zowghi, ``Gender disparity in the governance of software engineering conferences,'' in \emph{2019 IEEE/ACM 2nd International Workshop on Gender Equality in Software Engineering (GE)}.\hskip 1em plus 0.5em minus 0.4em\relax IEEE, 2019, pp. 21--24.

\bibitem{moldovan2024diversity}
V.~Moldovan and S.~Motogna, ``Diversity of se conferences,'' in \emph{Proceedings of the 5th ACM/IEEE Workshop on Gender Equality, Diversity, and Inclusion in Software Engineering}, 2024, pp. 47--54.

\bibitem{catolino2019gender}
G.~Catolino, F.~Palomba, D.~A. Tamburri, A.~Serebrenik, and F.~Ferrucci, ``Gender diversity and women in software teams: How do they affect community smells?'' in \emph{2019 IEEE/ACM 41st International Conference on Software Engineering: Software Engineering in Society (ICSE-SEIS)}.\hskip 1em plus 0.5em minus 0.4em\relax IEEE, 2019, pp. 11--20.

\bibitem{mathew2018finding}
G.~Mathew, A.~Agrawal, and T.~Menzies, ``Finding trends in software research,'' \emph{IEEE Transactions on Software Engineering}, vol.~49, no.~4, pp. 1397--1410, 2018.

\bibitem{chang2021job}
P.-C. Chang, H.~Rui, and T.~Wu, ``Job autonomy and career commitment: A moderated mediation model of job crafting and sense of calling,'' \emph{Sage Open}, vol.~11, no.~1, p. 21582440211004167, 2021.

\bibitem{esteves2017crafting}
T.~Esteves and M.~P. Lopes, ``Crafting a calling: The mediating role of calling between challenging job demands and turnover intention,'' \emph{Journal of Career Development}, vol.~44, no.~1, pp. 34--48, 2017.

\bibitem{replication_package}
\BIBentryALTinterwordspacing
``Replication package,'' 2024. [Online]. Available: \url{https://anonymous.4open.science/r/Female_Researchers-7656/}
\BIBentrySTDinterwordspacing

\bibitem{narayanan2023diversity}
A.~S. Narayanan, D.~Vagavolu, N.~A. Day, and M.~Nagappan, ``Diversity in software engineering conferences and journals,'' \emph{arXiv preprint arXiv:2310.16132}, 2023.

\bibitem{santana2021scientific}
T.~S. Santana and A.~H. Braga, ``A scientific research overview at the brazilian computer society congress: a feminine perspective.'' \emph{LAWCC@ CLEI}, pp. 93--104, 2021.

\bibitem{hosseini2021gender}
M.~Hosseini and S.~Sharifzad, ``Gender disparity in publication records: a qualitative study of women researchers in computing and engineering,'' \emph{Research Integrity and Peer Review}, vol.~6, pp. 1--14, 2021.

\bibitem{rodriguez2021perceived}
G.~Rodr{\'\i}guez-P{\'e}rez, R.~Nadri, and M.~Nagappan, ``Perceived diversity in software engineering: a systematic literature review,'' \emph{Empirical Software Engineering}, vol.~26, pp. 1--38, 2021.

\bibitem{frachtenberg2022underrepresentation}
E.~Frachtenberg and R.~D. Kaner, ``Underrepresentation of women in computer systems research,'' \emph{Plos one}, vol.~17, no.~4, p. e0266439, 2022.

\bibitem{freire2021measuring}
A.~Freire, L.~Porcaro, and E.~G{\'o}mez, ``Measuring diversity of artificial intelligence conferences,'' in \emph{Artificial Intelligence Diversity, Belonging, Equity, and Inclusion}.\hskip 1em plus 0.5em minus 0.4em\relax PMLR, 2021, pp. 39--50.

\bibitem{gomez2024diversity}
E.~Gomez, L.~Porcaro, P.~Frau~Amar, and J.~Vinagre, ``Diversity in artificial intelligence conferences,'' Joint Research Centre, Tech. Rep., 2024.

\bibitem{stathoulopoulos2019gender}
K.~Stathoulopoulos and J.~C. Mateos-Garcia, ``Gender diversity in ai research,'' \emph{Available at SSRN 3428240}, 2019.

\bibitem{CRedit}
\BIBentryALTinterwordspacing
NISO, ``{Origins of CRedit}.'' [Online]. Available: \url{https://credit.niso.org/origins/}
\BIBentrySTDinterwordspacing

\bibitem{allen2019can}
L.~Allen, A.~O’Connell, and V.~Kiermer, ``How can we ensure visibility and diversity in research contributions? how the contributor role taxonomy (credit) is helping the shift from authorship to contributorship,'' \emph{Learned Publishing}, vol.~32, no.~1, pp. 71--74, 2019.

\bibitem{niso2022ansi}
N.~C.~W. Group \emph{et~al.}, ``Ansi/niso z39. 104-2022, credit, contributor roles taxonomy.[s. l.],'' \emph{Baltimore, MD: National Information Standards Organization}, 2022.

\bibitem{zhu2021kill}
Q.~Zhu, A.~Zaidman, and A.~Panichella, ``How to kill them all: An exploratory study on the impact of code observability on mutation testing,'' \emph{Journal of Systems and Software}, vol. 173, p. 110864, 2021.

\bibitem{sagdic2024taxonomy}
\BIBentryALTinterwordspacing
E.~Sagdic, A.~Bayram, and M.~R. Islam, ``On the taxonomy of developers' discussion topics with chatgpt,'' in \emph{Proceedings of the 21st International Conference on Mining Software Repositories}, ser. MSR '24.\hskip 1em plus 0.5em minus 0.4em\relax New York, NY, USA: Association for Computing Machinery, 2024, p. 197–201. [Online]. Available: \url{https://doi.org/10.1145/3643991.3645080}
\BIBentrySTDinterwordspacing

\bibitem{JSS_Scopes}
\BIBentryALTinterwordspacing
JSS, ``Aims and scopes.'' [Online]. Available: \url{https://www.sciencedirect.com/journal/journal-of-systems-and-software/about/aims-and-scope}
\BIBentrySTDinterwordspacing

\bibitem{roebuck2019organizational}
A.~Roebuck, A.~Thomas, and B.~Biermeier-Hanson, ``Organizational culture mitigates lower ratings of female supervisors,'' \emph{Journal of Leadership \& Organizational Studies}, vol.~26, no.~4, pp. 454--464, 2019.

\bibitem{hengel2017publishing}
E.~Hengel, ``Publishing while female. are women held to higher standards? evidence from peer review.'' 2017.

\bibitem{lindahl2021importance}
J.~Lindahl, C.~Colliander, and R.~Danell, ``The importance of collaboration and supervisor behaviour for gender differences in doctoral student performance and early career development,'' \emph{Studies in Higher Education}, vol.~46, no.~12, pp. 2808--2831, 2021.

\bibitem{eagly2013female}
A.~H. Eagly, L.~Gartzia, and L.~L. Carli, ``Female advantage: revisited,'' 2013.

\bibitem{korabik1989should}
K.~Korabik and R.~Ayman, ``Should women managers have to act like men?'' \emph{Journal of Management Development}, vol.~8, no.~6, pp. 23--32, 1989.

\bibitem{kohl2021challenges}
K.~Kohl and R.~Prikladnicki, ``Challenges women in software engineering leadership roles face: A qualitative study,'' \emph{arXiv preprint arXiv:2104.13982}, 2021.

\bibitem{guzman2024mind}
E.~Guzm{\'a}n, R.~A.-L. Fischer, and J.~Kok, ``Mind the gap: gender, micro-inequities and barriers in software development,'' \emph{Empirical Software Engineering}, vol.~29, no.~1, p.~17, 2024.

\bibitem{andreessen2011software}
M.~Andreessen, ``Why software is eating the world,'' \emph{The Wall Street Journal}, vol.~8, p.~20, 2011.

\bibitem{foolLargestCompanies}
L.~Daly, ``The largest companies by market cap in 2024,'' \url{https://www.fool.com/research/largest-companies-by-market-cap/}, 2024.

\bibitem{mckinseyEconomicPotential}
\BIBentryALTinterwordspacing
M.~. Company, ``The economic potential of generative ai: The next productivity frontier,'' 2023. [Online]. Available: \url{https://www.mckinsey.com/capabilities/mckinsey-digital/our-insights/the-economic-potential-of-generative-ai-the-next-productivity-frontier#introduction}
\BIBentrySTDinterwordspacing

\end{thebibliography}
}
\end{document}